\documentclass[symmetry,article,moreauthors,pdflatex,10pt,a4paper]{Definitions/mdpi} 
\preto{\abstractkeywords}{\nolinenumbers}
\firstpage{1} 
\allowdisplaybreaks

\def\al{\alpha}
\def\be{\beta}
\def\ga{\gamma}
\def\de{\delta}

\def\th{\theta}

\def\si{\sigma}

\def\ch{\chi}
\def\ps{\psi}
\def\om{\omega}

\def\cH{{\cal H}}
\def\cl{{\cal L}}

\def\fr#1#2{{{#1}\over{#2}}}
\def\frac#1#2{{\textstyle{{#1}\over{#2}}}}
\def\half{{\textstyle{1\over 2}}}
\def\ol{\overline}
\def\prt{\partial}

\def\lsim{\mathrel{\rlap{\lower4pt\hbox{\hskip1pt$\sim$}}
    \raise1pt\hbox{$<$}}}
\def\gsim{\mathrel{\rlap{\lower4pt\hbox{\hskip1pt$\sim$}}
    \raise1pt\hbox{$>$}}}

\def\nn{\nonumber}

\def\vev#1{\langle {#1}\rangle}

\newcommand{\beq}{\begin{equation}}
\newcommand{\eeq}{\end{equation}}
\newcommand{\bea}{\begin{eqnarray}}
\newcommand{\eea}{\end{eqnarray}}
\newcommand{\Refe}[1]{Ref.~\cite{#1}}
\newcommand{\rf}[1]{(\ref{#1})}

\newcommand{\Eq}[1]{Equation (\ref{#1})}

\newcommand{\Tab}[1]{Table~\ref{#1}}
\newcommand{\Sec}[1]{Section~\ref{#1}}
\newcommand{\Subsec}[1]{Subsection~\ref{#1}}

\def\psb{\ol\ps{}}
\def\mbf#1{\boldsymbol #1}

\def\pvec{\mbf p}
\def\gavec{\mbf\ga}

\def\Q{\mathcal Q}
\def\S{\mathcal S}
\def\P{\mathcal P}

\def\Qhat{\widehat\Q}

\def\codt{\cos{\om_\oplus T_\oplus}}
\def\sodt{\sin{\om_\oplus T_\oplus}}
\def\ctodt{\cos{2\om_\oplus T_\oplus}}
\def\stodt{\sin{2\om_\oplus T_\oplus}}
\def\cthodt{\cos{3\om_\oplus T_\oplus}}
\def\sthodt{\sin{3\om_\oplus T_\oplus}}

\def\cmtemplate#1#2#3#4{{#1}^{#3}_{#4}}

\def\ctemplate#1#2#3#4{{#1}^{(#2)#3}_{#4}}

\def\bcf#1#2{\ctemplate{b}{#1}{#2}{F}}

\def\cmtemplate#1#2#3#4{{#1}^{#3}_{#4}}

\def\acmw#1#2#3{\cmtemplate{a}{#1}{#2}{{#3}}}
\def\bcmw#1#2#3{\cmtemplate{b}{#1}{#2}{{#3}}}
\def\ccmw#1#2#3{\cmtemplate{c}{#1}{#2}{{#3}}}
\def\dcmw#1#2#3{\cmtemplate{d}{#1}{#2}{{#3}}}
\def\ecmw#1#2#3{\cmtemplate{e}{#1}{#2}{{#3}}}

\def\gcmw#1#2#3{\cmtemplate{g}{#1}{#2}{{#3}}}
\def\Hcmw#1#2#3{\cmtemplate{H}{#1}{#2}{{#3}}}

\def\ctemplate#1#2#3#4{{#1}^{(#2)#3}_{#4}}
\def\mcw#1#2#3{\ctemplate{m}{#1}{#2}{{#3}}}

\def\acw#1#2#3{\ctemplate{a}{#1}{#2}{{#3}}}
\def\bcw#1#2#3{\ctemplate{b}{#1}{#2}{{#3}}}
\def\ccw#1#2#3{\ctemplate{c}{#1}{#2}{{#3}}}
\def\dcw#1#2#3{\ctemplate{d}{#1}{#2}{{#3}}}
\def\ecw#1#2#3{\ctemplate{e}{#1}{#2}{{#3}}}

\def\gcw#1#2#3{\ctemplate{g}{#1}{#2}{{#3}}}
\def\Hcw#1#2#3{\ctemplate{H}{#1}{#2}{{#3}}}

\def\mcfw#1#2#3{\ctemplate{m}{#1}{#2}{F{,#3}}}

\def\acfw#1#2#3{\ctemplate{a}{#1}{#2}{F{,#3}}}
\def\bcfw#1#2#3{\ctemplate{b}{#1}{#2}{F{,#3}}}
\def\ccfw#1#2#3{\ctemplate{c}{#1}{#2}{F{,#3}}}
\def\dcfw#1#2#3{\ctemplate{d}{#1}{#2}{F{,#3}}}
\def\ecfw#1#2#3{\ctemplate{e}{#1}{#2}{F{,#3}}}

\def\gcfw#1#2#3{\ctemplate{g}{#1}{#2}{F{,#3}}}
\def\Hcfw#1#2#3{\ctemplate{H}{#1}{#2}{F{,#3}}}

\def\ab{{\al\be}}

\def\mab{{\mu\al\be}}

\def\m{m_\ps}
\def\mw{m_w}

\def\cthodt{\cos{3\om_\oplus T_\oplus}}
\def\sthodt{\sin{3\om_\oplus T_\oplus}}

\def\atw#1#2{{\widetilde a}_{#1}^{#2}}
\def\btw#1#2{{\widetilde b}_{#1}^{#2}}
\def\bftw#1#2{{\widetilde b}_{F,#1}^{#2}}
\def\mftw#1#2{{\widetilde m}_{F,#1}^{#2}}
\def\atws#1#2{{\widetilde a}_{#1}^{*#2}}
\def\btws#1#2{{\widetilde b}_{#1}^{*#2}}
\def\bftws#1#2{{\widetilde b}_{F,#1}^{*#2}}
\def\mftws#1#2{{\widetilde m}_{F,#1}^{*#2}}

\def\bptw#1#2{{\widetilde b}_{#1}^{\prime #2}}

\def\ctw#1#2{{\widetilde c}_{#1}^{#2}}
\def\bptws#1#2{{\widetilde b}_{#1}^{\prime *#2}}

\def\ctws#1#2{{\widetilde c}_{#1}^{*#2}}

\def\vos{\mathrel{\rlap{\lower0pt\hbox{\hskip0.5pt{$\scriptstyle s$}}}
    \raise2pt\hbox{$\scriptstyle \neg$}}}

 \usepackage{environ}
\NewEnviron{myequation}{%
\begin{equation}
\scalebox{0.88}{$\BODY$}
\end{equation}}

\Title{Penning-Trap Searches for Lorentz and CPT Violation}

\Author{Yunhua Ding $^{1, 2,}$*, Teague D. Olewiler $^{2}$ and Mohammad Farhan Rawnak $^{2}$}

\AuthorNames{{Yunhua Ding}}

\address{$^{1}$ \quad W.M. Keck Science Department,
Claremont McKenna, Pitzer, and Scripps Colleges,
Claremont, CA 91711, USA; yding@kecksci.claremont.edu (Y.D.)\\
$^{2}$ \quad Department of Physics,
Gettysburg College,
Gettysburg, PA 17325, USA; olewte01@gettysburg.edu(T.D.O.);
rawnmo01@gettysburg.edu(M.F.R.)}

\corres{Correspondence: yding@kecksci.claremont.edu}

\abstract{
An overview of recent progress on testing Lorentz and CPT symmetry 
using Penning traps is presented.
The theory of quantum electrodynamics with Lorentz-violating 
operators of mass dimensions up to six is summarized. 
Dominant shifts in the cyclotron and anomaly frequencies of the confined particles 
and antiparticles due to Lorentz and CPT violation are derived.
Existing results of the comparisons of charge-to-mass ratios and magnetic moments 
involving protons, antiprotons, electrons, and positrons are used to
constrain various coefficients for Lorentz violation.
}

\keyword{Lorentz and CPT violation; Penning trap; Standard-Model Extension}

\begin{document}

\section{Introduction}
\label{Introduction}

Precision experiments involving Penning traps
have in recent years achieved impressive sensitivities to properties 
of fundamental particles. 
For example, 
the magnetic moment of electrons has been measured 
to a record precision of 0.28 
ppt~\cite{08ha}.
The high precision offered by Penning-trap experiments 
provides excellent opportunities to test fundamental symmetries.
This includes the Lorentz symmetry,
one of the foundations of both general relativity and the Standard Model of particle physics.
It has been recently shown that tiny violations of Lorentz symmetry
could naturally arise in a fundamental theory that unifies  
gravity with quantum physics at the Plank scale $M_P \sim 10^{19}$ GeV,
such as string 
theory~\cite{89ks,91kp}.
Since in any effective field theory,
violations of CPT symmetry also break Lorentz 
symmetry~\cite{ck,ck1,owg},
testing Lorentz symmetry also includes CPT tests.
In recent years, 
searches for Lorentz and CPT violation in precision experiments 
have been performed across many subfields of physics
\cite{dt},
including Penning-trap experiments. 
Here,
we provide an overview of the recent progress on testing Lorentz and CPT violation
in Penning-trap experiments measuring charge-to-mass ratios and magnetic moments 
of protons, antiprotons, electrons, and positrons.  
  
In the context of effective field theory,
the comprehensive framework that describes 
all possible Lorentz violation is the Standard-Model Extension 
(SME)~\cite{ck,ck1,akgrav}.
The Lagrange density of the SME is constructed from general relativity 
and the Standard Model by adding all possible Lorentz-violating terms.
Each of such terms is formed from a coordinate-independent 
contraction of a general Lorentz-violating operator with a corresponding coefficient.
The subset of the SME with operators of mass dimensions $d\leq 4$
is called the minimal SME,
which is power-counting renormalizable.
For the nonminimal SME, 
it restricts attention to operators of mass dimensions $d>4$,
which is viewed to produce higher-order effects.
Study of the nonminimal SME serves as a basis for further 
investigations of many aspects of Lorentz and CPT violation,
such the causality and 
stability~\cite{akrl,causality},
Lorentz-violating models in 
supersymmetry~\cite{susy},
noncommutative Lorentz-violating quantum 
electrodynamics~\cite{chklo,ncqft,ncqft2},
and the underlying pseudo-Riemann-Finsler 
geometry~\cite{finsler, finsler1,finsler2, 19ek}.

For Penning-trap experiments measuring charge-to-mass ratios
and magnetic moments of confined particles or antiparticles,
both the minimal and nonminimal SME can produce various measurable 
Lorentz- and CPT-violating effects via 
shifts in the cyclotron and anomaly frequencies.
These effects in general can depend on sidereal time 
and differ between particles and antiparticles.
In the minimal SME,
Refs.~\cite{bkr97,bkr98} present the first theoretical analysis 
to study Lorentz and CPT violation in Penning traps. 
An extension to the nonminimal SME by including 
Lorentz-violating operators of mass dimensions up to six
was recently made in 
\Refe{16dk},
in which analysis of the magnetic moment comparisons 
between particles and antiparticles using Penning traps was also performed.
A similar application to charge-to-ratio comparisons is presented 
in~\Refe{20dr}.
For the effects arising from sidereal variations due to the Earth's rotation,
the related discussions are given in 
~\Refe{mi99, 07ha, 19ding}.

In this work,
we provide an overview of recent progress on searching 
for Lorentz- and CPT-violating signals using Penning traps
and provide the most updated constraints 
on the coefficients for Lorentz violation that are relevant to these experiments.  
The results provided in this work are complementary to 
these from the studies of Lorentz and CPT violation 
in experiments involving measurements of the 
muon's anomalous magnetic moment 
~\cite{muon,gkv14},
the spectroscopic analysis of hydrogen, antihydrogen, and other related 
systems~\cite{kv15},
and clock 
comparisons~\cite{kv18}.

This work is organized as follows. 
We start in \Sec{theory} with the related theory,
where we present in \Subsec{Lagrange density} 
the theory of Lorentz-violating quantum electrodynamics with operators 
of mass dimensions up to six.
The perturbative energy shifts to the confined particles or antiparticles are obtained 
in \Subsec{perturbative energy shifts} using perturbation theory. 
\Subsec{cyclotron and anomaly frequencies} gives the corresponding shifts 
in the cyclotron and anomaly frequencies due to Lorentz and CPT violation. 
The discussion of sidereal variations and rotation matrices 
are given in \Subsec{sidereal variations}. 
Then we turn in \Sec{experiments} to experimental applications to various Penning-trap experiments 
and present the constraints on the coefficients for Lorentz violation. 
The applications to charge-to-mass ratio comparisons are treated in 
\Subsec{the charge-to-mass ratios},
with \Subsec{the proton sector} focusing on the proton sector, 
and \Subsec{the electron sector} discussing the electron sector,
respectively. 
The resulting constraints on the coefficients for Lorentz violation 
from the reported experimental results are summarized in 
\Tab{cons-p}, \Tab{cons-e}, and \Tab{cons-e-uw}.
\Subsec{the g factors and magnetic moments} discusses the applications to
magnetic moment comparisons,
with \Subsec{the proton sector 1} for the proton sector,
and  \Subsec{the electron sector 1} for the electron sector,
respectively. 
The corresponding limits on the coefficients for Lorentz violation 
are listed in \Tab{pgconstraints} and \Tab{epconstraints}.
The summary of this work is given in \Sec{summary}.
Finally, 
Appendix \ref{transformations} lists the explicit expressions of the transformations 
of the relevant coefficients into different frames.
Throughout this work,
we follow the same notation used in Refs.~\cite{16dk, 20dr}
and adopt natural units with $\hbar =c = 1$.

\section{Theory}
\label{theory}

In this section, 
we summarize the theory developed in~\Refe{16dk}
of Lorentz-violating quantum electrodynamics 
with operators of mass dimension $d\leq 6$
and derive the energy shifts for particles and antiparticles
confined in Penning traps due to Lorentz and CPT violation.

\subsection{Lagrange density}
\label{Lagrange density}

For a single Dirac fermion field $\ps$ of charge $q$ and mass $\m$,
the general Lorentz-violating Lagrange density $\cl_\ps$ 
can be constructed by adding a general Lorentz-violating operator $\Qhat$ 
to the conventional Lagrange density,
\bea
\cl_\ps =
\half \psb (\ga^\mu i D_\mu - \m + \Qhat) \ps + {\rm H.c.}, 
\label{fermlag}
\eea
where $D_\mu = (\prt_\mu + i q A_\mu)$ is the covariant derivative 
given by the minimal coupling with $A_\mu$ being the electromagnetic four-potential. 
H.c. means Hermitian conjugate.
The general Lorentz-violating operator $\Qhat$ in the Lagrange density \rf{fermlag} 
is a $4\times 4$ spinor matrix 
that contains terms formed by the contraction of a generic coefficient for Lorentz violation, 
the covariant derivative $iD_\mu$,
the antisymmetric electromagnetic field tensor 
$F_\ab \equiv \prt_\al A_\be - \prt_\be A_\al$,
and one of the 16 Dirac matrix bases. 
For example, 
one of the dimension-five operators involving the $F$-type coefficients for Lorentz violation
takes the form $\bcf 5 \mab F_\ab \ga_5 \ga_\mu $.
For mass dimension $d\leq 6$,
a full list of the relevant coefficients for Lorentz violation 
and their properties are given in Table I in \Refe{16dk}.
Note that the hermiticity of the Lagrange density \rf{fermlag} indicates that 
the operator $\Qhat$ satisfies the condition $\Qhat = \ga_0 \Qhat^\dag\ga_0$.
In the free-fermion limit where $A_\al = 0$,
the explicit expression of the Lagrange density \rf{fermlag} 
at arbitrary mass dimension has been studied in 
Ref.~\cite{km13}.
For the interaction case where $A_\al \neq 0$,
\Refe{16dk} 
developed a theory for operators of mass dimensions up to six.
An extension of the theory to include operators of arbitrary mass dimension
was recently given in
Ref.~\cite{19kl}.
Similar analysis has also been performed for other SME sectors, 
including these for
photon~\cite{km09},
neutrino~\cite{km12}
and 
gravity~\cite{nonmingrav}.

Due to the existence of the general operator $\Qhat$ in the Lagrange density \rf{fermlag},
the conventional Dirac equation for a fermion in electromagnetic fields is modified to
\bea
\label{moddirac}
(p \cdot \ga - \m + \Qhat ) \ps = 0 ,
\eea
where we have chosen the momentum space for convenience,
with the identification $p_\al  \leftrightarrow iD_\al$.
Given the fact that no Lorentz-violating signals have been observed so far,
any such signal must be tiny compared to the energy scale of the system of interest.
Therefore,
we can treat the corrections due to Lorentz and CPT violation 
to the conventional Hamiltonian as perturbative 
and apply perturbation theory to obtain the dominant shifts 
to the energy levels of the confined particles and antiparticles.
From \Eq{moddirac},
the exact Hamiltonian $\cH$ can be defined as
\bea
\cH \ps \equiv p^0 \ps = \ga_0 (\pvec \cdot \gavec + \m - \Qhat)\ps=(\cH_0 + \de \cH)\ps,
\eea
where $p^0$ is the exact energy of the system of interest,
including all contributions from Lorentz and CPT violation,
$\cH_0$ is the conventional Hamiltonian for a fermion in an electromagnetic field,
and $\de \cH=-\ga_0 \Qhat$ is the exact perturbative Hamiltonian.

To construct $\de \cH$,
we note that operator $\Qhat$ in general contains terms that are of powers of $p^0$, 
which corresponds to the exact Hamiltonian $\cH$ itself. 
In certain simple cases,
one can apply a field redefinition to remove the additional time derivatives
and adopt the standard procedure involving time translation on wave functions
to obtain the exact perturbative Hamiltonian 
$\de \cH$~\cite{bkr98}.
In more general cases, 
it is challenging to directly construct $\de \cH$ due to the existence of powers of time derivatives.
However,
we notice that any contributions to $\de\cH$ due to the exact Hamiltonian $\cH$ 
are at second or higher orders in the coefficients for Lorentz violation.
Therefore, 
to obtain the leading-order results,
one can apply the following substitution 
\cite{km12, 16dk},
\bea
\de\cH \approx -\ga_0 \Qhat |_{p^0\to E_0},
\label{perturbH}
\eea
where $E_0$ is the unperturbative eigenvalue,
which can be obtained by solving the conventional Dirac equation for a fermion in 
an electromagnetic field.

\subsection{Perturbative energy shifts}
\label{perturbative energy shifts}

With the perturbative Hamiltonian $\de \cH$ determined by \Eq{perturbH},
the shifts in the energy levels of a confined particle can be obtained 
by applying perturbation theory, 
\bea
\de E_{n, \pm}= \vev{\ch_{n, \pm}|\de\cH|\ch_{n, \pm}},
\label{deE}
\eea
where $\ch_{n, \pm}$ are the unperturbative stationary eigenstates 
with $n$ being the level number and $\pm$ denoting the spin
for a positive-energy fermion.
$\de E_{n, \pm}$ are the perturbative corrections to the energy levels 
due to Lorentz and CPT violation.

To derive the expressions for $\de E_{n, \pm}$, 
we can take an idealized Penning trap
where a constant uniform magnetic field is applied to confine the particle's radial motion
and a quadrupole electric field provides the axial confinement.
The dominant effects in the unperturbative energy levels 
are due to the interactions between the confined particle and the magnetic field.
The quadruple electric field generates effects suppressed by a factor of $E/B\approx 10^{-5}$ 
in natural units for a typical Penning trap field configuration with 
$E \approx 20$~kV/m and $B \approx 5$~T.
Therefore, 
the leading-order results in $\de E_{n, \pm}$ can be obtained by  
further idealizing the trap as a pure uniform magnetic field 
in which a quantum fermion moves.
Following the above discussion,
for a spin-1/2 fermion,
the leading-order perturbative energy shifts due to Lorentz and CPT violation are found to 
be
\cite{20dr} 
\bea
\label{linear-en}
\de E_{n, \pm 1}^w 
&=&
\atw w 0 
\mp \si \btw w 3
- \mftw w 3 B
\pm \si \bftw w {33} B 
+ \big(
\pm \si \bptw w {3} 
- m_w [\ctw w {00} + (\ctw w {11} + \ctw w {22})_{s}]
\big)
\fr{(2n+1\mp \si)|qB|}{2 m_w^2}
\nn\\
&&
\hskip -3pt
+ 
\big(
\mp \si (\btw w {311} + \btw w {322})
- \fr{1}{m_w} (\ctw w {11} + \ctw w {22})_{\vos}
\big) 
\fr{(2n+1)|qB|}{2},
\eea
where $\si$ is the charge sign of the fermion, given by $q\equiv \si |q|$.
The subscript $w$ indicates the fermion species.
For example, 
for electrons and protons, 
$w=e$ and $w=p$, 
respectively.
The tilde coefficients
are defined by 
\bea
\label{tild}
\atw w 0
&=&
\acmw 3 0 w
- \mw \ccmw 4 {00} w
- \mw \ecmw 4 0 w
+ \mw^2 \mcw 5 {00} w
+ \mw^2 \acw 5 {000} w
- \mw^3 \ccw 6 {0000} w
- \mw^3 \ecw 6 {000} w,
\nn\\
\btw w 3
&=&
\bcmw 3 3 w
+ \Hcmw 3 {12} w
- \mw \dcmw 4 {30} w
- \mw \gcmw 4 {120} w
+ \mw^2 \bcw 5 {300} w
+ \mw^2 \Hcw 5 {1200} w
- \mw^3 \dcw 6 {3000} w
- \mw^3 \gcw 6 {12000} w,
\nn\\
\mftw w 3
&=&
\mcfw 5 {12} w
+ \acfw 5 {012} w
- \mw \ccfw 6 {0012} w
- \mw \ecfw 6 {012} w,
\nn\\
\bftw w {33}
&=&
\bcfw 5 {312} w
+ \Hcfw 5 {1212} w
- \mw \dcfw 6 {3012} w
- \mw \gcfw 6 {12012} w ,
\nn\\
\bptw w {3}
&=&
b_{w}^{3} + m_w (g_{w}^{120} - g_{w}^{012} + g_{w}^{021}) 
- m_w^2 b_{w}^{(5)300} 
- 2 m_w^2 (H_{w}^{(5)1200} - H_{w}^{(5)0102} + H_{w}^{(5)0201})
\nn \\
&&
+ 2 m_{w}^3 d_{w}^{(6)3000}  
+ 3 m_w^3 (g_{w}^{(6)12000}-g_{w}^{(6)01002}+g_{w}^{(6)02001}) ,
\nn\\
\ctw w {00}
&=&
c_{w}^{00} - m_w m_{w}^{(5)00} - 2 m_w a_{w}^{(5)000} 
+ 3 m_w^2 c_{w}^{(6)0000} + 2 m_w^2 e_{w}^{(6)000},
\eea
and the ``11+22" types of tilde coefficients are defined by
\bea 
\label{jj}
(\ctw w {jj})_{s}
& = &
c_{w}^{jj} 
- 2 m_w a_{w}^{(5)j0j} 
+ 3 m_w^2 c_{w}^{(6)j00j}  ,
\nn\\ 
(\ctw w {jj} )_{\vos}
& = &
- m_w a_{w}^{(5)0jj} 
- m_w m_{w}^{(5)jj} 
+ 3 m_w^2 c_{w}^{(6)00jj} 
+ 3 m_w^2 e_{w}^{(6)0jj} ,
\nn\\
\btw w {3jj} 
& = &
b_{w}^{(5)3jj} + H_{w}^{(5)12jj} 
- 3 m_{w} d_{w}^{(6)30jj} 
- 3 m_{w} g_{w}^{(6)120jj} ,
\eea
with $j=1$ or $2$ only.  
The subscripts $s$ and 
$\mathrel{\rlap{\lower0pt\hbox{\hskip0.5pt{$s$}}}\raise2pt\hbox{$\neg$}}$
in the $\ctw w {jj}$ tilde coefficients in the energy shifts \rf{linear-en}  
show that $(\ctw w {jj})_{s}$ give both spin-dependent
and spin-independent energy shifts,
while $(\ctw w {jj} )_{\vos}$ produce only spin-independent ones,
as evident from the corresponding proportional factors 
$2n+1\mp \si$ and $2n+1$.
We note that the tilde coefficients 
$\atw w 0$,
$\btw w 3$,
$\mftw w 3$ ,
and $\bftw w {33}$ in the energy shifts \rf{linear-en} 
produce effects that are independent of the level number $n$.

Following a similar analysis, 
the corresponding leading-order shifts to the energy levels of antifermions
can be determined by using
\bea
\label{deEs}
\de E^c_{n,\pm}= \vev{\ch^c_{n,\pm}|\de\cH^c |\ch^c_{n,\pm}} ,
\eea
where $\ch^c_{n,\pm}$ are the positive-energy antifermion eigenstates,
obtained by applying charge conjugation 
on the negative-energy fermion solutions $\ch_{n,\pm}$.
$\de\cH^c$ is the antifermion perturbative Hamiltonian,
which can be obtained from $\de\cH$ 
by reversing the charge sign $\sigma$ and changing the signs for 
all CPT-odd coefficients in operator $\Qhat$.
Using \Eq{deEs},
the antifermion results are given by 
\cite{20dr}
\bea
\label{linear-ens}
\de E_{n, \pm 1}^{\ol{w}} 
&=&
\ - \atws w 0 
\pm \si \btws w 3
- \mftws w 3 B
\mp \si \bftws w {33} B 
+ \big(
\mp \si \bptws w {3} 
- m_w [\ctw w {00} + (\ctw w {11} + \ctw w {22})_{s}]
\big)
\fr{(2n+1\mp \si)|qB|}{2 m_w^2}
\nn \\
&&
+ 
\big(
\pm \si (\btws w {311} + \btws w {322})
- \fr{1}{m_w} (\ctw w {11} + \ctw w {22})_{\vos}
\big) 
\fr{(2n+1)|qB|}{2},
\eea
where the tilde quantities with a star subscript are given by 
\bea
\label{tilds}
\atws w 0
&=&
\acmw 3 0 w
+ \mw \ccmw 4 {00} w
- \mw \ecmw 4 0 w
- \mw^2 \mcw 5 {00} w
+ \mw^2 \acw 5 {000} w
+ \mw^3 \ccw 6 {0000} w
- \mw^3 \ecw 6 {000} w,
\nn\\
\btws w 3
&=&
\bcmw 3 3 w
- \Hcmw 3 {12} w
+ \mw \dcmw 4 {30} w
- \mw \gcmw 4 {120} w
+ \mw^2 \bcw 5 {300} w
- \mw^2 \Hcw 5 {1200} w
+ \mw^3 \dcw 6 {3000} w
- \mw^3 \gcw 6 {12000} w,
\nn\\
\mftws w 3
&=&
\mcfw 5 {12} w
- \acfw 5 {012} w
- \mw \ccfw 6 {0012} w
+ \mw \ecfw 6 {012} w,
\nn\\
\bftws w {33}
&=&
\bcfw 5 {312} w
- \Hcfw 5 {1212} w
+ \mw \dcfw 6 {3012} w
- \mw \gcfw 6 {12012} w ,
\nn\\
\bptws w {3}
&=&
b_{w}^{3} + m_w (g_{w}^{120} - g_{w}^{012} + g_{w}^{021}) 
- m_w^2 b_{w}^{(5)300} 
+ 2 m_w^2 (H_{w}^{(5)1200} - H_{w}^{(5)0102} + H_{w}^{(5)0201})
\nn \\
&&
- 2 m_{w}^3 d_{w}^{(6)3000}  
+ 3 m_w^3 (g_{w}^{(6)12000}-g_{w}^{(6)01002}+g_{w}^{(6)02001}) ,
\nn\\
\ctws w {00}
&=&
c_{w}^{00} - m_w m_{w}^{(5)00} + 2 m_w a_{w}^{(5)000} 
+ 3 m_w^2 c_{w}^{(6)0000} - 2 m_w^2 e_{w}^{(6)000},
\eea
and the corresponding ``11+22" starred tilde quantities are defined by
\bea 
\label{jjs}
(\ctws w {jj})_{s}
& = &
c_{w}^{jj} 
+ 2 m_w a_{w}^{(5)j0j} 
+ 3 m_w^2 c_{w}^{(6)j00j}  ,
\nn\\ 
(\ctws w {jj} )_{\vos}
& = &
m_w a_{w}^{(5)0jj} 
- m_w m_{w}^{(5)jj} 
+ 3 m_w^2 c_{w}^{(6)00jj} 
- 3 m_w^2 e_{w}^{(6)0jj} ,
\nn\\
\btws w {3jj} 
& = &
b_{w}^{(5)3jj} - H_{w}^{(5)12jj} 
+ 3 m_{w} d_{w}^{(6)30jj} 
- 3 m_{w} g_{w}^{(6)120jj} .
\eea
In the result~\rf{linear-ens},
the charge sign $\si$ is understood to be reversed for the antifermion.
Comparing the fermion and antifermion energy shifts results 
\rf{linear-en} and \rf{linear-ens},
$\de E_{n, \pm 1}^{\ol{w}}$ can also been obtained 
from $\de E_{n, \pm 1}^{w}$  
by reversing the charge sign $\si$,
the spin orientation,
and the signs of all CPT-odd coefficients for Lorentz violation,
as expected. 

We remark in passing that 
the rotation properties of the coefficients for Lorentz violation appearing in 
results \rf{linear-en} and \rf{linear-ens}
are represented by their indices. 
For example,
the index pair ``12" on the right sides of the definitions \rf{tild} and \rf{tilds}
is antisymmetric.
This implies that these coefficients for Lorentz violation 
transform like a single ``3" index under rotations,
while coefficients with an index ``0" or an index pair ``00" are invariant under rotations. 
The cylindrical symmetry of the Penning trap is correctly reflected 
by the fact that the results \rf{linear-en} and \rf{linear-ens} only depend on 
index ``0", ``3", and ``11+22".

\subsection{Cyclotron and anomaly frequencies}
\label{cyclotron and anomaly frequencies}

The primary observables of interest in a Penning-trap experiment are frequencies. 
Two key frequencies are the cyclotron frequency $\nu_c \equiv \om_c/2\pi$ and
the Larmor spin-precession frequency $\nu_L \equiv \om_L/2\pi$.
The difference of the two frequencies gives 
the anomaly frequency $\nu_L-\nu_c = \nu_a \equiv  \om_a/2\pi$
\cite{geo}.
For a confined fermion of flavor $w$ in a Penning trap,
the cyclotron and anomaly frequencies are defined as the
energy difference between the following energy levels 
\cite{bkr98, 16dk},
\bea
\label{fermi}
\om_c^{w} 
=
 E_{1,\si}^{w}- E_{0,\si}^{w} ,
\qquad
 \om_a^{w} 
=
 E_{0,-\si}^{w}- E_{1,\si}^{w}.
\eea
For an antifermion of flavor $\ol{w}$, 
the corresponding definitions for the cyclotron and anomaly frequencies are given by 
\cite{bkr98, 16dk}
\bea
\label{antifermi}
\om_c^{\ol{w}}
=
 E_{1,\si}^{\ol{w}}- E_{0,\si}^{\ol{w}} ,
\qquad
 \om_a^{\ol{w}} 
=
 E_{0,-\si}^{\ol{w}}- E_{1,\si}^{\ol{w}},
\eea
with the understanding that the charge signs $\si$ in definitions (\ref{antifermi})
are reversed compared to these in definitions (\ref{fermi}).

In a Lorentz-invariant scenario,
the charge-to-mass ratio and the $g$ factor
of a particle or an antiparticle confined in a Penning trap 
with a magnetic field strength $B$ 
are related to the above cyclotron and anomaly frequencies by
\bea
\label{qm}
\dfrac{|q|}{m}   = \dfrac{\om_c}{B} ,
\eea  
 and
\bea
\fr {g}{2} = \fr {\om_L}{\om_c}
=
1+ \fr {\om_a}{\om_c},
\label{ratio}
\eea
respectively. 
For Penning-trap experiments comparing  
the charge-to-mass ratios or the $g$ factors between a particle of species $w$ 
and its corresponding antiparticle $\ol{w}$,
the CPT theorem guarantees that the following differences must be zero, 
\bea
\label{ratio-li}
\dfrac{(|q|/m)_{\ol{w}}}{(|q|/m)_{w}} - 1
=
\dfrac{\om_c^{\ol{w}}}{\om_c^{w}} - 1 
=0,
\eea
and 
\beq
\half (g_w - g_{\ol w})
=
\fr {\om_a^w}{\om_c^w} - \fr {\om_a^{\ol w}}{\om_c^{\ol w}} 
=0,
\label{omdifference}
\eeq 
where in \Eq{ratio-li},
for simplicity,
we have assumed the same magnetic field is used in the comparison.
When different magnetic fields are used,
the expression can be easily obtained by substituting 
$\om_c^{\ol{w}}/\om_c^{w}$ in the middle term of \Eq{ratio-li} by
$(\om_c^{\ol{w}}/B_{\ol{w}})(\om_c^{w}/B_w)$,
where $B_w$ and $B_{\ol{w}}$ are the strengths of the magnetic fields 
used to confine the particle and antiparticle,
respectively.  
However, 
this assumption is not required to derive \Eq{omdifference}
as the ratios $\om_a^w/\om_c^w$ and $\om_a^{\ol{w}}/\om_c^{\ol{w}}$
don't depend on the magnetic field used in the trap.

However,
in the presence of Lorentz and CPT violation,
the picture changes dramatically as
both the cyclotron and anomaly frequencies for fermions and antifermions 
can be shifted due to the Lorentz- and CPT-violating corrections to the energy levels,
\bea
\label{cyclnochange}
\de \om_c^{w} 
=
\de E_{1,\si}^{w}- \de E_{0,\si}^{w} ,
\qquad
\de \om_a^{w} 
=
\de E_{0,-\si}^{w}- \de E_{1,\si}^{w},
\nn\\
\de \om_c^{\ol w} 
=
\de E_{1,\si}^{\ol w}-\de E_{0,\si}^{\ol w} 
\qquad
\de \om_a^{\ol w}
=
\de E_{0,-\si}^{\ol w}-\de E_{1,\si}^{\ol w}.
\eea
Applying energy shift results \rf{linear-en} and \rf{linear-ens},
the corrections to the cyclotron frequencies for a fermion and antifermion 
are found to be 
\cite{20dr}
\bea
\label{wcshift}
\de \omega_c^{w}
&=&
\left(\dfrac{1}{m_w^2} \bptw w {3} 
- \dfrac{1}{m_w} (\ctw w {00} + \ctw w {11} + \ctw w {22})
- (\btw w {311} + \btw w {322})\right) eB ,
\nn\\
\de \omega_c^{\ol{w}}
&=&
\left(- \dfrac{1}{m_w^2} \bptws w {3} 
- \dfrac{1}{m_w} (\ctws w {00} + \ctws w {11} + \ctws w {22})
+ (\btws w {311} + \btws w {322})\right)eB ,
\eea
where the tilde and starred tilde coefficients are given 
by definitions \rf{tild} and \rf{tilds}.
The $\ctw w {jj}$ and $\ctws w {jj}$ tilde coefficients with $j$ taking values of $1$ or $2$ 
are the sum of the two quantities given in definitions \rf{jj} and~\rf{jjs},
\bea
\ctw w {jj} 
&=&
(\ctw w {jj})_{s} + (\ctw w {jj} )_{\vos}
\nn\\
&=&
c_{w}^{jj} 
- 2 m_w a_{w}^{(5)j0j} 
+ 3 m_w^2  c_{w}^{(6)j00j} 
- m_w a_{w}^{(5)0jj} 
- m_w m_{w}^{(5)jj} 
+ 3 m_w^2 c_{w}^{(6)00jj} 
+ 3 m_w^2 e_{w}^{(6)0jj}  ,
\nn\\
\ctws w {jj} 
&=&
(\ctws w {jj})_{s} + (\ctws w {jj} )_{\vos}
\nn\\
&=&
c_{w}^{jj} 
+ 2 m_w a_{w}^{(5)j0j} 
+ 3 m_w^2  c_{w}^{(6)j00j} 
+ m_w a_{w}^{(5)0jj} 
- m_w m_{w}^{(5)jj} 
+ 3 m_w^2 c_{w}^{(6)00jj} 
- 3 m_w^2 e_{w}^{(6)0jj}  .
\eea
For the shifts in the anomaly frequencies,
the results are found to be 
\cite{16dk}
\bea
\label{washift}
\de \om_a^{w} 
&=&
 2 \btw w 3 - 2 \bftw w {33} B ,
\nn\\
\de \om_a^{\ol w}
&=&
- 2 \btws w 3 + 2 \bftws w {33} B ,
\eea
where various tilde and starred tilde coefficients are given 
by definitions \rf{tild}, \rf{tilds}, \rf{jj} and \rf{jjs}.
We note in passing that the shifts in the cyclotron and anomaly frequencies 
for an antifermion in results \rf{wcshift} and \rf{washift} 
reveals that they can be obtained from these for the fermion by changing the signs
of all the basic coefficients for Lorentz violation that control CPT-odd effects,
as might be expected. 

Result \rf{wcshift} shows that 
the cyclotron frequency shifts due to Lorentz and CPT violation 
for a fermion are different from these for its corresponding antifermion. 
The same conclusion holds for the anomaly frequency shifts from the result \rf{washift}.
This implies that in the presence of Lorentz and CPT violation,
the differences \rf{qm} and \rf{ratio-li} do not vanish in general.
For the charge-to-mass ratio comparisons, 
the result becomes
\bea
\label{ratio-lv}
\dfrac{(|q|/m)_{\ol{w}}}{(|q|/m)_{w}} - 1
\longleftrightarrow
\dfrac{\om_c^{\ol{w}}}{\om_c^{w}} - 1
=
\dfrac{\de \om_c^{\ol{w}} -  \de \om_c^{w}} {\om_c^{w}},
\eea
where the Lorentz- and CPT-invariant pieces 
in the measured cyclotron frequencies are exactly canceled
by the CPT theorem if the same magnetic field is used.
The notation $\longleftrightarrow$ indicates the correspondence between 
the experimental interpreted charge-to-mass ratio comparison
and the measured frequency difference 
$\om_c^{\ol{w}}/\om_c^{w} - 1$.
For the $g$ factor comparison between a fermion and antifermion,
the related ratio comes to be
\bea
\label{gratio-lv}
\half (g_w - g_{\ol w})
\longleftrightarrow
\fr {\om_a^w}{\om_c^w} - \fr {\om_a^{\ol w}}{\om_c^{\ol w}} =
\fr {\de\om_a^w}{\om_c^w} - \fr {\de\om_a^{\ol w}} {\om_c^{\ol w}},
\eea 
where again all Lorentz- and CPT-invariant contributions are canceled out 
on the right side. 
We note that in the above expression \rf{gratio-lv},
we only keep shifts in the anomaly frequencies $\de \om_a^w$ and $\de \om_a^{\ol{w}}$ 
due to Lorentz and CPT violation,
as contributions from the cyclotron frequencies $\de \om_c^w$ and $\de {\om_c^{\ol {w}}}$ 
are suppressed by factors of $eB/m_w^2$,
as evident from results \rf{wcshift} and \rf{washift}.
Even for a comparatively large magnetic field of $B \approx 5$ T in a Penning trap,
these factors are at orders of $eB/m_e^2 \approx 10^{-9}$ for electrons or positrons,
and $eB/m_p^2 \approx 10^{-16}$ for protons or antiprotons,
which can be ignored in \Eq{gratio-lv}.

\subsection{Sidereal variations}
\label{sidereal variations}

The cyclotron and anomaly frequency shifts  \rf{wcshift} and \rf{washift},
which also appear in comparisons~\rf{ratio-lv} and \rf{gratio-lv},
are obtained in a particular apparatus frame $x^a\equiv (x^1,x^2,x^3)$,
where the positive $\hat x_3$ axis is aligned with the applied magnetic field in the trap.
However, 
as Earth rotates about its axis, 
this apparatus frame is not inertial. 
The standard canonical frame that is adopted in the literature 
to compare results from different experiments searching for 
Lorentz and CPT violation is called the Sun-centered frame 
$X^J\equiv (X,Y,Z)$~\cite{sunframe0,sunframe}.
In this frame,
the $Z$ axis is defined to be aligned along the Earth's rotation axis,
the $X$ axis points from the Earth to the Sun, 
and the $Y$ axis completes a right-handed coordinate system.
The time origin of this coordinate system 
is chosen to be at the vernal equinox 2000.
The coefficients for Lorentz violation in this frame 
are assumed to be constants in time and 
space~\cite{ck,akgrav}.

To explicitly express the relationship of the coefficients for Lorentz violation
between the Sun-centered frame $X^J\equiv (X,Y,Z)$ 
and the apparatus frame $x^a\equiv (x^1,x^2,x^3)$,
it is convenient to introduce a third frame called the standard laboratory frame
$x^j \equiv (x,y,z)$,
with the $z$ axis pointing to the local zenith,
the $x$ axis aligned with the local south,
and the $y$ axis completing a right-handed coordinate system. 
To relate the coordinates of these three frames,
we define two rotation matrices
$R^{aj}$ and $R^{jJ}$ \cite{sunframe0,sunframe},
with $R^{aj}$ connecting $x^j\equiv (x,y,z)$ to $x^a\equiv (x^1,x^2,x^3)$ 
by $x^a = R^{aj} x^j$, 
and $R^{jJ}$ relating $X^J\equiv (X, Y, Z)$  
to $(x,y,z)$ by $x^{j} = R^{jJ} X^J$.
The expressions for these two rotation matrices are given by
\beq
R^{jJ}=\left(
\begin{array}{ccc}
\cos\ch\cos\om_\oplus T_\oplus
&
\cos\ch\sin\om_\oplus T_\oplus
&
-\sin\ch
\\
-\sin\om_\oplus T_\oplus
&
\cos\om_\oplus T_\oplus
&
0
\\
\sin\ch\cos\om_\oplus T_\oplus
&
\sin\ch\sin\om_\oplus T_\oplus
&
\cos\ch
\end{array}
\right),
\label{rot}
\eeq
and
\bea
R^{aj} &=&
\left(
\begin{array}{ccc}
\cos\ga &\sin\ga  &0\\
-\sin\ga &\cos\ga &0\\
0 &0 &1\\
\end{array} 
\right)
\times
\left(
\begin{array}{ccc}
\cos\be &0 &-\sin\be\\
0 &1 &0\\
\sin\be &0 &\cos\be\\
\end{array} 
\right)
\times
\left(
\begin{array}{ccc}
\cos\al &\sin\al &0\\
-\sin\al &\cos\al &0\\
0 &0 &1\\
\end{array} 
\right),
\label{euler}
\eea
where $\om_\oplus \approx 2\pi/(23{\rm ~h} ~56{\rm ~min})$
is the sidereal frequency of the Earth's rotation,
$T_\oplus$ denotes the local sidereal time,
$\ch$ specifies the colatitude of the laboratory,
and
$(\al, \be, \ga)$ are the Euler angles
in the convenient ``$y$-convention" of the rotation.
The coordinates of the apparatus frame
and the Sun-centered frame can then be related by
using the following expression,
\bea
\label{axis}
x^{a} = R^{aj} x^j = R^{aj} R^{jJ} x^J .
\eea
The relationship between the coefficients for Lorentz violation
in these two frames can also be derived from this result. 
We note that in the case that a vertical upward magnetic field is used in the trap,
the Euler angles become $(\al, \be, \ga)=(0,0,0)$ and the rotation matrix $R^{aj}$ 
reduces to the identity matrix.

The above transformation \rf{axis} shows that 
in general the coefficients for Lorentz violation and 
thus the cyclotron and anomaly frequency shifts  \rf{wcshift} and \rf{washift}
determined in the apparatus frame
depend on the sidereal time and the geometric location of the laboratory.
As a result, 
signals observed in Earth-based experiments,
including the above comparisons \rf{ratio-lv} and \rf{gratio-lv},
can oscillate at harmonics of the Earth's sidereal frequency $\om_\oplus$,
with the amplitudes depending on the laboratory colatitude~$\chi$.
To explicitly illustrate this,
here we consider a Penning-trap experiment located at colatitude $\ch$ 
and assume the magnetic field is aligned with the $z$ axis.
We give two explicit examples of the transformation of coefficients for Lorentz violation
using \Eq{axis}.
We first focus on the single-index laboratory-frame coefficient $\bptw w 3$,
which appears in the shifts of the cyclotron frequency~\rf{wcshift}. 
Applying the transformation \rf{axis} and taking $(\al, \be, \ga)=(0,0,0)$ imply 
\beq
\label{trans-one}
\bptw w 3 =
 \cos\om_\oplus T_\oplus \bptw w X \sin\ch
+ \sin\om_\oplus T_\oplus \bptw w Y \sin\ch 
+\bptw w Z \cos\ch .
\eeq
This result relates the tilde coefficients $\bptw w 3$ observed 
in the noninertial apparatus frame to the constant coefficients  
$\btw w J$ with $J=X,Y,Z$
in the canonical inertial Sun-centered frame.
The expression contains terms proportional to the first harmonic in the Earth's sidereal frequency 
and a constant. 
The colatitude dependence is evident from the factors $\sin\ch$ and $\cos\chi$ appearing above. 

A slightly more complicated result can also be obtained for the sum of the two-index tilde quantities 
$\ctw w {11} + \ctw w {22}$ in an analogous fashion.
Applying the rotation \rf{axis} for each index with $R^{aj}$ being the identity matrix yields 
\bea
\label{trans-two}
\ctw w {11} + \ctw w {22} 
&=& 
\ctodt  
\big(  - \half (\ctw w {XX} -  \ctw w {YY}) \sin^2 \ch \big)
+ \stodt
(  -  \ctw w {(XY)} \sin^2 \ch)
\nn \\
&& 
+  \codt  
(  - \ctw w {(XZ)}  \sin 2\ch )
+ \sodt
(  - \ctw w {(YZ)}  \sin 2\ch  )
\nn \\
&&
+ 
\frac{1}{4} (\ctw w {XX} + \ctw w {YY})(3+\cos 2\ch) + \ctw w {ZZ} \sin^2\ch ,
\eea 
where a parenthes on two indices of the coefficients
indicates symmetrization with a factor of 1/2.
For example, 
$\ctw w {(XY)} = (\ctw w {XY} + \ctw w {YX} )/2$.
It is revealed from the result \rf{trans-two} 
that the sum of the tilde coefficients $\ctw w {11} + \ctw w {22}$
in the apparatus frame depends on six independent quantities 
$\ctw w {(JK)}$ with $J,K=X,Y, Z$ in the Sun-centered frame,
producing up to second harmonics in the sidereal frequency of the Earth's rotation. 
For both examples \rf{trans-one} and  \rf{trans-two}, 
if the applied magnetic field points along a generic direction,
trigonometric functions of the Euler angles
($\al$, $\be$, $\ga$)
appear as well.

At the end of this subsection, 
we note that the revolution of the Earth about the Sun 
can generate additional types of time variations for the coefficients for Lorentz violation, 
such as the boost of the Earth relative to the Sun $\be_\oplus \approx 10^{-4}$,
and the boost of the laboratory due to the Earth's rotation $\be_L \approx 10^{-6}$. 
However, 
as studied in the 
literature~\cite{gkv14,kv15,ca04,he08,space},
such boost effects are suppressed by one or more powers of their boost factors
$\be_\oplus$ and $\be_L$
compared to these from rotations.
Therefore,
we treat them as negligible effects in the present work.

\section{Experiments}
\label{experiments}

In this section, 
we analyze existing Penning-trap experiments 
that compare the charge-to-mass ratios 
and the $g$ factors between protons, antiprotons, 
electrons, and positrons. 
We use reported results for the comparisons \rf{ratio-lv} and \rf{gratio-lv}
from  these experiments to constrain the relevant Sun-centered frame
tilde coefficients for Lorentz violation.

\subsection{The charge-to-mass ratios}
\label{the charge-to-mass ratios}

For charge-to-mass ratio comparisons
between a particle and its corresponding antiparticle, 
the result \rf{ratio-lv} implies that
the relevant tilde coefficients for Lorentz violation are
$\bptw w {3}$, 
$\ctw w {11} + \ctw w {22}$,
$\btw w {311} + \btw w {322}$,
$\bptws w {3}$, 
$\ctws w {11} + \ctws w {22}$,
and
$\btws w {311} + \btws w {322}$
in the apparatus frame,
given in the result \rf{wcshift}.
The transformations of these tilde coefficients 
into the Sun-centered frame depend on the field configuration 
in the trap. 
For a typical Penning-trap experiment, 
the magnetic field is
oriented either horizontally or vertically.
\Refe{20dr}
presents a complete list of transformation results for these tilde coefficients 
for the above two field orientations. 
Here, 
for completeness,
we include them in Appendix \ref{transformations}. 
From these transformation results, 
it shows that for a given fermion of species $w$, 
the relevant quantities in the Sun-centered frame that 
are related to the charge-to-mass comparisons in a Penning trap are  
the following 54 independent tilde coefficients,
$\bptw w {J}$,
$\ctw w {(JK)}$,
$\btw w {J(KL)}$,
$\bptws w {J}$,
$\ctws w {(JK)}$,
and 
$\btws w {J(KL)}$. 
In the following subsections, 
we apply the reported precisions for the charge-to-mass ratio comparisons from
Penning-trap experiments to set bounds on the relevant tilde coefficients for Lorentz violation.

\subsubsection{The proton sector}
\label{the proton sector}

We start the analysis with the charge-to-mass comparisons 
between protons and antiprotons.
In a Penning-trap experiment located at CERN by the ATRAP collaboration,
Gabrielse and his group achieved a precision of 90 ppt 
for the proton-antiproton charge-to-mass ratio 
comparison~\cite{ga99}. 
The experiment used a trap with a vertical uniform magnetic field 
$B = 5.85$~T.
Recently, 
another Penning-trap experiment at CERN by the BASE collaboration
led by Ulmer improved the comparison to the record sensitivity of 69 
ppt~\cite{ul15},
by applying a horizontal magnetic field $B = 1.946$ T 
which is oriented $60^\circ$ east of north.
For the measurement of the charge-to-mass ratio of a proton,
both experiments used a trapped hydrogen ion (H$^-$) as a proxy for the proton 
to eliminate systematic shifts caused by polarity switching of the trapping voltages.
The charge-to-mass ratio comparison between an antiproton and a proton
is then related to that between an antiproton and a hydrogen ion by
\beq
\label{ratioH1}
\dfrac{(|q|/m)_{\bar{p}}}{(|q|/m)_{p}} - 1
=
\dfrac{(|q|/m)_{\bar{p}}}{R (|q|/m)_{\rm{H}^-}} - 1
\longleftrightarrow
\dfrac{\de \om_c^{\bar{p}} - R \de \om_c^{\rm{H}^-}} {R \om_c^{\rm{H}^-}} ,
\eeq
where $R=m_{\rm{H}^-}/m_p = 1.001089218754$
is the ratio of the mass between a hydrogen ion and a 
proton~\cite{ul15},
and $\om_c^{\rm{H}^-}$ is the cyclotron
frequency for the hydrogen ion,
with $\de \om_c^{\rm{H}^-}$ being the corresponding shifts. 
To obtain $\de \om_c^{\rm{H}^-}$,
one can take $w=\rm{H}^-$ in the expression~\rf{wcshift}
and the related tilde coefficients for Lorentz violation 
become the effective ones for a hydrogen ion.
Expressing these effective coefficients in terms of the 
corresponding fundamental coefficients for the hydrogen ion constituents,
which are the coefficients for electrons and protons,
is challenging due to nonperturbative issues including binding effects 
in the composite hydrogen ion. 
However, 
an approximation to these coefficient relations can be obtained by 
treating the wave function of the hydrogen ion as a product of 
the wave functions of a proton and two electrons.
Applying perturbation theory at the lowest order
and ignoring the related binding energies,
the cyclotron frequency shifts $\de \om_c^{\rm{H}^-}$ of the hydrogen ion 
due to Lorentz and CPT violation can then be approximated as the sum of these for its constituents,
$\de \om_{c}^{\rm{H}^-} \approx \de \om_{c}^{p} + 2 \de \om_{c}^{e^-}$.
Substituting this into the result \rf{ratioH1} yields
\bea
\label{ratioH2}
\dfrac{(|q|/m)_{\bar{p}}}{(|q|/m)_{p}} - 1
\longleftrightarrow
\dfrac{\de \om_c^{\bar{p}} - R \de \om_c^{p} -2R  \de \om_c^{e^-}} {R \om_c^{\rm{H}^-}} .
\eea
As shown from the above result, 
the choice of using a hydrogen ion as a proxy for the proton 
in the Penning trap provides sensitivities not only to the
coefficients for Lorentz violation in the proton sector,
but also introduces additional sensitivities to these for electrons. 
Putting the above discussion together, 
the related Sun-centered frame tilde coefficients for Lorentz violation 
that are sensitive to Penning-trap experiments comparing 
the charge-to-mass ratios between protons and antiprotons 
are the following 81 independent tilde quantities,
$\bptw p {J}$,
$\ctw p {(JK)}$,
$\btw p {J(KL)}$,
$\bptws p {J}$,
$\ctws p {(JK)}$,
$\btws p {J(KL)}$,
$\bptw e {J}$,
$\ctw e {(JK)}$,
and
$\btw e {J(KL)}$.
The published results for the comparison \rf{ratio-lv}
from both the ATRAP and the BASE experiments
can be adopted to set bounds on these tilde coefficients for Lorentz violation.

For the ATRAP experiment, 
the reported precision was obtained by analyzing the measurements of the 
cyclotron frequencies in a time-averaged way,
so any oscillating effects in the difference~\rf{ratioH2} averaged out.
This implies that only the constant terms that appear in the transformation results in 
the first half of Appendix~\ref{transformations} can be constrained using the published precision. 
Applying expression \rf{ratioH2} by taking the reported precision of 90 ppt for
$(|q|/m)_{\bar{p}}/(|q|/m)_{p} - 1$
and identifying $\om_c^{\rm{H}^-} =  2\pi \times 89.3$ MHz 
given in the ATRAP experiment,
the following limit can be obtained,
\beq
\label{limit-atrap}
| \de \om_c^{\bar{p}} - 1.001 \de \om_c^{p} -2.002  \de \om_c^{e^-}|_{\rm{const}} 
\lsim 3.33 \times 10^{-26}\ \rm{GeV},
\eeq
where the subscript ``const" indicates that only the constant terms
in the transformations results are relevant to the limit. 
However,
future sidereal-variation analysis of the measurements 
of the cyclotron frequencies can provide constraints on 
the non-constant terms that appear in the harmonics 
in the transformation results.  

For the experiment carried out by the BASE collaboration,
since the magnetic field was oriented at
$60^\circ$ east of north,
both of the transformation matrices \rf{rot} and \rf{euler} are required to relate  
the tilde coefficients for Lorentz violation in the apparatus frame 
to these in the Sun-centered frame.
For a general horizontal magnetic field with an angle~$\th$ 
measured from the local south in the counterclockwise direction,
the corresponding Euler angles are found to be $(\al, \be, \ga) =  (\th, \pi/2, 0)$.
The second half of Appendix \ref{transformations} 
lists the full expressions of the transformations for the related 
tilde coefficients for Lorentz violation.
The BASE experiment analyzed the data of the charge-to-mass ratio comparisons
to search for both time-averaged effects and 
sidereal variations in the first harmonics of the Earth's rotation frequency.
Therefore,
the reported results can be taken to set bounds on not only the constant terms 
but also on the terms proportional to the first harmonics in the transformation results
given in Appendix \ref{transformations}. 
Using the reported 69 ppt for the time-averaged precision
and 720 ppt for the limit of the first harmonic amplitude for the comparison \rf{ratioH2}, 
and taking $\om_c^{\rm{H}^-} =  2\pi \times 29.635$ MHz for the BASE experiment,
the following limits are obtained,
\beq
\label{limit-base}
| \de \om_c^{\bar{p}} - 1.001 \de \om_c^{p} -2.002  \de \om_c^{e^-}|_{\rm{const}} 
\lsim 8.46 \times 10^{-27}\ \rm{GeV}
\eeq
and 
\beq
\label{limit-base-1st}
| \de \om_c^{\bar{p}} - 1.001 \de \om_c^{p} -2.002  \de \om_c^{e^-}|_{\rm{1st}} 
\lsim 8.83 \times 10^{-26}\ \rm{GeV},
\eeq
where the subscript ``const" in the limit \rf{limit-base} 
has similar meaning as the one in \rf{limit-atrap},
while the subscript ``1st" in the limit \rf{limit-base-1st} specifies only
the amplitude of the first harmonics in the sidereal variation.   

The limit \rf{limit-atrap} for the ATRAP experiment  
and limits \rf{limit-base} and \rf{limit-base-1st} for the BASE experiment 
set constraints on a combination of the Sun-centered frame tilde coefficients  
$\bptw p {J}$,
$\ctw p {(JK)}$,
$\btw p {J(KL)}$,
$\bptws p {J}$,
$\ctws p {(JK)}$,
$\btws p {J(KL)}$,
$\bptw e {J}$,
$\ctw e {(JK)}$,
and
$\btw e {J(KL)}$.
These constraints can be obtained by substituting result \rf{wcshift} in limit~\rf{limit-atrap}
and applying the corresponding transformations given in Appendix \ref{transformations} 
with $\ch =43.8^\circ$ for the ATRAP experiment,
and substituting result \rf{wcshift} in both limits \rf{limit-base} and \rf{limit-base-1st}
and identifying $\th=2 \pi/3$ and $\ch =43.8^\circ$ in the transformation expressions
in Appendix \ref{transformations} for the BASE experiment. 
To get some intuition about the scope of the constraints
on the individual component of the tilde coefficients,
we can take a common practice that is adopted in many subfields 
searching for Lorentz and CPT 
violation~\cite{dt} 
which assumes that only one individual tilde coefficient is nonzero at a time.
From the ATRAP limit~\rf{limit-atrap}, 
constraints on 27 independent tilde coefficients for Lorentz violation
are obtained,
while from the BASE limits \rf{limit-base} and \rf{limit-base-1st},
a total of 69 independent tilde coefficients for Lorentz violation 
are constrained. 
We summarize in \Tab{cons-p} and  \Tab{cons-e} the constraints  
in the proton sector and the electron sector,
respectively.
In both tables, 
the first column lists the individual components, 
the second column presents the corresponding constraint 
on the modulus of the component,
the third column specifies the related experiment,
and the related reference is given in the final column. 
We note that some of the tilde coefficients for Lorentz violation 
are constrained by both the ATRAP and the BASE experiments.
To keep the results clean, 
we only keep the more stringent ones in both tables.

\renewcommand{\arraystretch}{1.4}
\begin{table}[H]
\centering
\caption{
\label{cons-p}
Constraints on tilde coefficients for Lorentz violation in the proton sector using
the charge-to-mass ratio comparisons from the ATRAP and the BASE experiments.}
\setlength{\tabcolsep}{5pt}
\begin{tabular}{llcc}
\hline
\hline																														Coefficient						&				Constraint				&	Experiment	&	Reference	\\	\hline
$	|	 \bptw p {Z}	|,	|	 \bptws p {Z}	|	$	&	$	<	1	\times 	10^{-10}	$	GeV	&	ATRAP	&	\cite{ga99}	\\	
$	|	 \ctw p {XX}	|,	|	 \ctws p {XX}	|	$	&	$	<	1	\times 	10^{-10}	$		&	ATRAP	&	\cite{ga99}	\\	
$	|	 \ctw p {YY}	|,	|	 \ctws p {YY}	|	$	&	$	<	1	\times 	10^{-10}	$		&	ATRAP	&	\cite{ga99}	\\	
$	|	 \ctw p {ZZ}	|,	|	 \ctws p {ZZ}	|	$	&	$	<	8	\times 	10^{-11}	$		&	BASE	&	\cite{ul15}	\\	
$	|	 \btw p {X(XZ)}	|,	|	 \btws p {X(XZ)}	|	$	&	$	<	2	\times 	10^{-10}	$	 GeV$^{-1}$	&	BASE	&	\cite{ul15}	\\	
$	|	 \btw p {Y(YZ)}	|,	|	 \btws p {Y(YZ)}	|	$	&	$	<	2	\times 	10^{-10}	$	 GeV$^{-1}$	&	BASE	&	\cite{ul15}	\\	
$	|	 \btw p {ZZZ}	|,	|	 \btws p {ZZZ}	|	$	&	$	<	2	\times 	10^{-10}	$	 GeV$^{-1}$	&	BASE	&	\cite{ul15}	\\	
$	|	 \btw p {ZXX}	|,	|	 \btws p {ZXX}	|	$	&	$	<	2	\times 	10^{-10}	$	 GeV$^{-1}$	&	ATRAP	&	\cite{ga99}	\\	
$	|	 \btw p {ZYY}	|,	|	 \btws p {ZYY}	|	$	&	$	<	2	\times 	10^{-10}	$	 GeV$^{-1}$	&	ATRAP	&	\cite{ga99}	\\	
																					
$	|	 \bptw p {X}	|,	|	 \bptws p {X}	|	$	&	$	<	7	\times 	10^{-10}	$	GeV	&	BASE	&	\cite{ul15}	\\	
$	|	 \bptw p {Y}	|,	|	 \bptws p {Y}	|	$	&	$	<	7	\times 	10^{-10}	$	GeV	&	BASE	&	\cite{ul15}	\\	
$	|	 \ctw p {(XZ)}	|,	|	 \ctws p {(XZ)}	|	$	&	$	<	1	\times 	10^{-9}	$		&	BASE	&	\cite{ul15}	\\	
$	|	 \ctw p {(YZ)}	|,	|	 \ctws p {(YZ)}	|	$	&	$	<	1	\times 	10^{-9}	$		&	BASE	&	\cite{ul15}	\\	
$	|	 \btw p {XXX}	|,	|	 \btws p {XXX}	|	$	&	$	<	2	\times 	10^{-9}	$	 GeV$^{-1}$	&	BASE	&	\cite{ul15}	\\	
$	|	 \btw p {X(XY)}	|,	|	 \btws p {X(XY)}	|	$	&	$	<	2	\times 	10^{-9}	$	 GeV$^{-1}$	&	BASE	&	\cite{ul15}	\\	
$	|	 \btw p {XYY}	|,	|	 \btws p {XYY}	|	$	&	$	<	1	\times 	10^{-9}	$	 GeV$^{-1}$	&	BASE	&	\cite{ul15}	\\	
$	|	 \btw p {XZZ}	|,	|	 \btws p {XZZ}	|	$	&	$	<	9	\times 	10^{-10}	$	 GeV$^{-1}$	&	BASE	&	\cite{ul15}	\\	
$	|	 \btw p {YXX}	|,	|	 \btws p {YXX}	|	$	&	$	<	1	\times 	10^{-9}	$	 GeV$^{-1}$	&	BASE	&	\cite{ul15}	\\	
$	|	 \btw p {Y(XY)}	|,	|	 \btws p {Y(XY)}	|	$	&	$	<	2	\times 	10^{-9}	$	 GeV$^{-1}$	&	BASE	&	\cite{ul15}	\\	
$	|	 \btw p {YYY}	|,	|	 \btws p {YYY}	|	$	&	$	<	2	\times 	10^{-9}	$	 GeV$^{-1}$	&	BASE	&	\cite{ul15}	\\	
$	|	 \btw p {YZZ}	|,	|	 \btws p {YZZ}	|	$	&	$	<	9	\times 	10^{-10}	$	 GeV$^{-1}$	&	BASE	&	\cite{ul15}	\\	
$	|	 \btw p {Z(XZ)}	|,	|	 \btws p {Z(XZ)}	|	$	&	$	<	3	\times 	10^{-9}	$	 GeV$^{-1}$	&	BASE	&	\cite{ul15}	\\	
$	|	 \btw p {Z(YZ)}	|,	|	 \btws p {Z(YZ)}	|	$	&	$	<	3	\times 	10^{-9}	$	 GeV$^{-1}$	&	BASE	&	\cite{ul15}	\\						
\hline
\hline
\end{tabular}
\end{table}

\renewcommand{\arraystretch}{1.4}
\begin{table}[H]
\centering
\caption{
\label{cons-e}
Constraints on tilde coefficients for Lorentz violation in the electron sector using
the charge-to-mass ratio comparisons from the ATRAP and the BASE experiments.}
\setlength{\tabcolsep}{5pt}
\begin{tabular}{llcc}
\hline
\hline																											Coefficient						&				Constraint				&	Experiment	&	Reference	\\	\hline			
$	|	 \bptw e {Z}	|				$	&	$	<	2	\times 	10^{-17}	$	GeV	&	ATRAP	&	\cite{ga99}	\\	
$	|	 \ctw e {XX}	|				$	&	$	<	3	\times 	10^{-14}	$		&	ATRAP	&	\cite{ga99}	\\	
$	|	 \ctw e {YY}	|				$	&	$	<	3	\times 	10^{-14}	$		&	ATRAP	&	\cite{ga99}	\\	
$	|	 \ctw e {ZZ}	|				$	&	$	<	2	\times 	10^{-14}	$		&	BASE	&	\cite{ul15}	\\	
$	|	 \btw e {X(XZ)}	|				$	&	$	<	1	\times 	10^{-10}	$	 GeV$^{-1}$	&	BASE	&	\cite{ul15}	\\	
$	|	 \btw e {Y(YZ)}	|				$	&	$	<	1	\times 	10^{-10}	$	 GeV$^{-1}$	&	BASE	&	\cite{ul15}	\\	
$	|	 \btw e {ZZZ}	|				$	&	$	<	1	\times 	10^{-10}	$	 GeV$^{-1}$	&	BASE	&	\cite{ul15}	\\	
$	|	 \btw e {ZXX}	|				$	&	$	<	9	\times 	10^{-11}	$	 GeV$^{-1}$	&	ATRAP	&	\cite{ga99}	\\	
$	|	 \btw e {ZYY}	|				$	&	$	<	9	\times 	10^{-11}	$	 GeV$^{-1}$	&	ATRAP	&	\cite{ga99}	\\	
																					
$	|	 \bptw e {X}	|				$	&	$	<	1	\times 	10^{-16}	$	GeV	&	BASE	&	\cite{ul15}	\\	
$	|	 \bptw e {Y}	|				$	&	$	<	1	\times 	10^{-16}	$	GeV	&	BASE	&	\cite{ul15}	\\	
$	|	 \ctw e {(XZ)}	|				$	&	$	<	3	\times 	10^{-13}	$		&	BASE	&	\cite{ul15}	\\	
$	|	 \ctw e {(YZ)}	|				$	&	$	<	3	\times 	10^{-13}	$		&	BASE	&	\cite{ul15}	\\	
$	|	 \btw e {XXX}	|				$	&	$	<	1	\times 	10^{-9}	$	 GeV$^{-1}$	&	BASE	&	\cite{ul15}	\\	
$	|	 \btw e {X(XY)}	|				$	&	$	<	9	\times 	10^{-10}	$	 GeV$^{-1}$	&	BASE	&	\cite{ul15}	\\	
$	|	 \btw e {XYY}	|				$	&	$	<	5	\times 	10^{-10}	$	 GeV$^{-1}$	&	BASE	&	\cite{ul15}	\\	
$	|	 \btw e {XZZ}	|				$	&	$	<	5	\times 	10^{-10}	$	 GeV$^{-1}$	&	BASE	&	\cite{ul15}	\\	
$	|	 \btw e {YXX}	|				$	&	$	<	5	\times 	10^{-10}	$	 GeV$^{-1}$	&	BASE	&	\cite{ul15}	\\	
$	|	 \btw e {Y(XY)}	|				$	&	$	<	9	\times 	10^{-10}	$	 GeV$^{-1}$	&	BASE	&	\cite{ul15}	\\	
$	|	 \btw e {YYY}	|				$	&	$	<	1	\times 	10^{-9}	$	 GeV$^{-1}$	&	BASE	&	\cite{ul15}	\\	
$	|	 \btw e {YZZ}	|				$	&	$	<	5	\times 	10^{-10}	$	 GeV$^{-1}$	&	BASE	&	\cite{ul15}	\\	
$	|	 \btw e {Z(XZ)}	|				$	&	$	<	2	\times 	10^{-9}	$	 GeV$^{-1}$	&	BASE	&	\cite{ul15}	\\	
$	|	 \btw e {Z(YZ)}	|				$	&	$	<	2	\times 	10^{-9}	$	 GeV$^{-1}$	&	BASE	&	\cite{ul15}	\\	 \hline
\hline
\end{tabular}
\end{table}

\subsubsection{The electron sector}
\label{the electron sector}

For the comparison of the charge-to-mass ratios between
an electron and a positron,
the current most accurate result was made in an experiment 
at the University of Washington,
with a precision at 130 ppb
\cite{81sc}. 
From the comparison \rf{ratio-lv}, 
this time-average result gives the following limit,
\beq
\label{limit-e-uw}
| \de \om_c^{e^-} -  \de \om_c^{e^+} |_{\rm{const}} 
\lsim 7.66 \times 10^{-20}\ \rm{GeV}.
\eeq
Following a similar analysis as the one used for the proton sector in the previous subsection,
one can use the colatitude and magnetic field that are relevant to this experiment
$\ch = 42.5^\circ$ and $B=5.1$ T upward,
together with the transformation expressions given in Appendix \ref{transformations}, 
to obtain the constraints on the tilde coefficients for Lorentz violation 
$\bptw e {J}$,
$\ctw e {(JK)}$,
$\btw e {J(KL)}$,
$\bptws e {J}$,
$\ctws e {(JK)}$,
and
$\btws e {J(KL)}$.
We summarize the results in \Tab{cons-e-uw},
which is organized in the same way as \Tab{cons-p} and \Tab{cons-e}. 
We note that some of the constraints of the tilde coefficients
from the electron-positron charge-to-mass ratio comparison 
are not comparable to these given in \Tab{cons-e} from the proton-antiproton comparison,
so we don't include them in \Tab{cons-e-uw}.
We also note that the Penning trap experiment at the University of Washington 
used a radioactive source of positrons that requires special precautions.
Efforts in applying a safe source and ensuring efficient positron accumulation 
are currently being made at both Harvard University and Northwestern University
\cite{15fo, 19gf},
providing great potential for improving the 
current bounds for the tilde coefficients listed in \Tab{cons-e} and \Tab{cons-e-uw}.

\renewcommand{\arraystretch}{1.4}
\begin{table}[H]
\centering
\caption{
\label{cons-e-uw}
Constraints on tilde coefficients for Lorentz violation in the electron sector using
the charge-to-mass ratio comparisons from the experiment at the University of Washington.}
\setlength{\tabcolsep}{5pt}
\begin{tabular}{llcc}
\hline
\hline																															Coefficient						&				Constraint				&	Experiment	&	Reference	\\	\hline
$	|	 \bptws e {Z}	|				$	&	$	<	9	\times 	10^{-11}	$	GeV	&	Washington	&	\cite{81sc}	\\	
$	|	 \ctws e {XX}	|				$	&	$	<	2	\times 	10^{-7}	$		&	Washington	&	\cite{81sc}	\\	
$	|	 \ctws e {YY}	|				$	&	$	<	2	\times 	10^{-7}	$		&	Washington	&	\cite{81sc}	\\	
$	|	 \ctws e {ZZ}	|				$	&	$	<	3	\times 	10^{-7}	$		&	Washington	&	\cite{81sc}	\\	
$	|	 \btws e {X(XZ)}	|				$	&	$	<	8	\times 	10^{-4}	$	 GeV$^{-1}$	&	Washington	&	\cite{81sc}	\\	
$	|	 \btws e {Y(YZ)}	|				$	&	$	<	8	\times 	10^{-4}	$	 GeV$^{-1}$	&	Washington	&	\cite{81sc}	\\	
$	|	 \btws e {ZZZ}	|				$	&	$	<	8	\times 	10^{-4}	$	 GeV$^{-1}$	&	Washington	&	\cite{81sc}	\\	
$	|	 \btws e {ZXX}	|				$	&	$	<	4	\times 	10^{-4}	$	 GeV$^{-1}$	&	Washington	&	\cite{81sc}	\\	
$	|	 \btws e {ZYY}	|				$	&	$	<	4	\times 	10^{-4}	$	 GeV$^{-1}$	&	Washington	&	\cite{81sc}	\\		\hline
\hline
\end{tabular}
\end{table}

\subsection{The $g$ factors and magnetic moments}
\label{the g factors and magnetic moments}

For the $g$ factor and magnetic moment comparisons 
between particles and antiparticles using Penning traps,
the result \rf{gratio-lv} implies that the relevant coefficients for Lorentz violation
in the apparatus frame are these given in \Eq{washift},
$\btw w 3$, $\btws w 3$, $\bftw w {33}$, and~$\bftws w {33}$.
For a Penning trap with a vertical or horizontal magnetic field,
the expressions of the transformation for these tilde coefficients 
into the Sun-centered frame are also given in Appendix \ref{transformations}.
The results show that there are 18 independent components of the tilde coefficients
for each fermion species $w$,
given by  
$\btw w J$, $\btws w J$, $\bftw w {(JK)}$, and~$\bftws w {(JK)}$.
In the subsection follows, 
we list the related Penning-trap experiments and 
use the reported comparison results to set bounds on
the corresponding tilde coefficients for Lorentz violation.

\subsubsection{The proton sector}
\label{the proton sector 1}

In the proton sector,
the current best measurements of the magnetic moments for both a 
proton and an antiproton were achieved by the BASE collaboration.
The proton magnetic moment measurement 
has a sensitivity of 0.3 ppb using a Penning trap located at Mainz
~\cite{17sc},
improving their previous best measurement~\cite{mo14}
by a factor of 11.
The antiproton magnetic moment measurement
was measured to a precision of 1.5 ppb 
with a similar Penning trap located at CERN
~\cite{17sm}.
Combining the reported 0.3 ppb and 1.5 ppb precisions 
for the time-averaged measurements,
and identifying $\om_c^{p}=2\pi\times 28.96$ MHz
and 
$\om_c^{\ol p}=2\pi\times 29.66$ MHz
for each experiment,
comparison \rf{gratio-lv} yields 
\bea
\label{base-g}
| \de\om_a^p - 0.98 \de\om_a^{\ol p} |_{\rm{const}} 
\lsim  9.53 \times 10^{-25}\ \rm{GeV},
\eea
where same subscript ``const" is used to specify 
only the constant terms in the transformation are relevant to the above limit.

The corresponding constraints for the combinations of the tilde coefficients 
in the Sun-centered frame can be obtained by applying the corresponding transformation 
results given in Appendix \ref{transformations} that are related to the specific field configuration in the trap 
and substituting  the numerical values of the laboratory quantities for both experiments in limit \rf{base-g}.
For the proton magnetic moment measurement at Mainz,
the colatitude is $\ch = 40.0^\circ$ 
and the magnetic field $B= 1.9$ T points 
$\th=18^\circ$ from local south in the counterclockwise direction
~\cite{17sc}.
For the experiment measuring the antiproton magnetic moment at CERN,
the trap is located at a slight different colatitude $\ch = 43.8^\circ$ 
and the magnetic field $B= 1.95$~T points 
$\th=120^\circ$ in the same convention as above
~\cite{17sm}.
Adopting the same assumption that only one tilde coefficient is nonzero at a time, 
the constraint on each independent tilde coefficient can be obtained. 
We summarize them in \Tab{pgconstraints} in the same fashion as before.

\renewcommand{\arraystretch}{1.4}
\begin{table}[H]
	\centering
	\caption{
	Constraints on tilde coefficients for Lorentz violation in the proton sector using
the $g$ factor comparison from the BASE experiments at Mainz and CERN.
	}\label{pgconstraints}
	\setlength{\tabcolsep}{5pt}
	\begin{tabular}{llcc}
\hline
\hline
		Coefficient			&			Constraint			&	Experiment	&	Reference	\\	\hline
$	|	\btw p Z	|	$	&	$	<	8\times 10^{-25}	{\rm ~GeV}	$	&	BASE	&	\cite{17sc, 17sm}	\\	
$	|	\btws p Z	|	$	&	$	<	1\times 10^{-24}	{\rm ~GeV}	$	&	BASE	&	\cite{17sc, 17sm}	\\	
$	|	\bftw p {XX} + \bftw p {YY}	|	$	&	$	<	4\times 10^{-9}	{\rm ~GeV}^{-1}	$	&	BASE	&	\cite{17sc, 17sm}	\\	
$	|	\bftw p {ZZ}	|	$	&	$	<	3\times 10^{-9}	{\rm ~GeV}^{-1}	$	&	BASE	&	\cite{17sc, 17sm}	\\	
$	|	\bftws p {XX} + \bftws p {YY}	|	$	&	$	<	3\times 10^{-9}	{\rm ~GeV}^{-1}	$	&	BASE	&	\cite{17sc, 17sm}	\\	
$	|	\bftws p {ZZ}	|	$	&	$	<	1\times 10^{-8}	{\rm ~GeV}^{-1}	$	&	BASE	&	\cite{17sc, 17sm}	\\	\hline
\hline	
	\end{tabular}
\end{table}

At the end of this subsection,
we point out that a sidereal-variation analysis of the anomaly frequencies 
for both protons and antiprotons is currently being carried out by the BASE collaboration.
This could in principle set bounds on other components of the tilde coefficients 
$\btw p J$, $\bftw p {(JK)}$, $\btws p J$ and $\bftws p {(JK)}$
that haven't been constrained before. 
The BASE collaboration is also developing a quantum logic readout system 
to allow more rapid measurements of the anomaly frequencies in the 
trap~\cite{chip},
which will offer excellent opportunities to perform a sidereal-variation analysis of the experimental data,
with great potential to set more stringent limits on the tilde coefficients for Lorentz violation.

\subsubsection{The electron sector}
\label{the electron sector 1}

In the electron sector, 
the comparison of the anomaly frequencies between electrons and positrons 
were performed in a Penning-trap experiment at the University of Washington,
with a precision of about
2~ppt~\cite{de99}.
The experimental data were analyzed in a time-averaged way 
to obtain a constraint of $b \lsim 50$ rad/s using the notation given in \Refe{de99}. 
Translating to the notation in this work yields 
\beq
\label{limit-uw}
| \de \om_a^{e^-} - \de \om_a^{e^+}|_{\rm{const}} 
\lsim 2.09 \times 10^{-23}\ \rm{GeV}.
\eeq
Taking the experimental quantities $\ch = 42.5^\circ$ and $B=5.85$ T upward,  
as well as the transformation presented in Appendix \ref{transformations},
the limit \rf{limit-uw} can be converted to constraints on the following Sun-centered frame
tilde coefficients,
$\btw e Z$, $\btws e Z$,
$\bftw e {XX} + \bftw e {YY}$,	
$\bftws e {XX} + \bftws e {YY}$,
$\bftw e {ZZ}$,
and $\bftws e {ZZ}$.

The measurement of the $g$ factor for electrons 
has reached a record precision of 0.28 ppt at Harvard University
~\cite{08ha}.
A sidereal-variation analysis of the anomaly frequencies was performed
at the frequencies of $\om_{\oplus}$ and $2\om_{\oplus}$,
yielding the same limit on the amplitudes 
of both the first and the second harmonics in the oscillation,
$|\de \om_a^e|_{\rm{1st/2nd}}  \lsim 2\pi \times 0.05$
Hz~\cite{07ha}.
Converting the results in natural units
gives the following limits,
\beq
\label{limit-1st-harvard}
| \de \om_a^{e^-}|_{\rm{1st}} 
\lsim 2.07 \times 10^{-25} \ \rm{GeV},
\eeq
and 
\beq
\label{limit-2nd-harvard}
| \de \om_a^{e^-}|_{\rm{2nd}} 
\lsim 2.07 \times 10^{-25}\ \rm{GeV}.
\eeq
Taking the magnetic field adopted in the experiment as $B=5.36$ T 
in the local upward direction
with the geometrical colatitude $\ch=47.6^\circ$
and applying the transformation results in Appendix \ref{transformations},
constraints on the following additional components of the tilde coefficients,
$\btw e X$, $\btw e Y$, 
$\bftw e {(XY)}$, $\bftw e {(XZ)}$,  $\bftw e {(YZ)}$,
and
$\bftw e {XX} - \bftw e {YY}$
are obtained.
The results from the University of Washington and Harvard University
are summarized in \Tab{epconstraints}.

\renewcommand{\arraystretch}{1.4}
\begin{table}[H]
	\centering
	\caption{
		Constraints on tilde coefficients for Lorentz violation in the electron sector using
the $g$ factor measurements from the experiments at the University of Washington and Harvard University.
	}\label{epconstraints}
	\setlength{\tabcolsep}{5pt}
	\begin{tabular}{llcc}
\hline
\hline
		Coefficient			&			Constraint			&	Experiment	&	Reference	\\	\hline
$	|	\btw e X	|	$	&	$	<	1\times 10^{-25}	{\rm ~GeV}	$	&	Harvard	&	\cite{07ha}	\\	
$	|	\btw e Y	|	$	&	$	<	1\times 10^{-25}	{\rm ~GeV}	$	&	Harvard	&	\cite{07ha}	\\	
$	|	\btw e Z	|	$	&	$	<	7\times 10^{-24}	{\rm ~GeV}	$	&	Washington	&	\cite{de99}	\\	
$	|	\btws e Z	|	$	&	$	<	7\times 10^{-24}	{\rm ~GeV}	$	&	Washington	&	\cite{de99}	\\	
$	|	\bftw e {XX} + \bftw e {YY}	|	$	&	$	<	2\times 10^{-8}	{\rm ~GeV}^{-1}	$	&	Washington	&	\cite{de99}	\\	
$	|	\bftw e {ZZ}	|	$	&	$	<	8\times 10^{-9}	{\rm ~GeV}^{-1}	$	&	Washington	&	\cite{de99}	\\	
$	|	\bftw e {(XY)} 	|	$	&	$	<	2\times 10^{-10}	{\rm ~GeV}^{-1}	$	&	Harvard	&	\cite{07ha}	\\	
$	|	\bftw e {(XZ)} 	|	$	&	$	<	1\times 10^{-10}	{\rm ~GeV}^{-1}	$	&	Harvard	&	\cite{07ha}	\\	
$	|	\bftw e {(YZ)} 	|	$	&	$	<	1\times 10^{-10}	{\rm ~GeV}^{-1}	$	&	Harvard	&	\cite{07ha}	\\	
$	|	\bftws e {XX} + \bftws e {YY}	|	$	&	$	<	2\times 10^{-8}	{\rm ~GeV}^{-1}	$	&	Washington	&	\cite{de99}	\\	
$	|	\bftws e {XX} - \bftws e {YY}	|	$	&	$	<	4\times 10^{-10}	{\rm ~GeV}^{-1}	$	&	Harvard	&	\cite{07ha}	\\	
$	|	\bftws e {ZZ}	|	$	&	$	<	8\times 10^{-9}	{\rm ~GeV}^{-1}	$	&	Washington	&	\cite{de99}	\\		
\hline
\hline
	\end{tabular}
\end{table}

The measurement of the magnetic moment for positrons
are currently under development at both Harvard University and Northwestern University
~\cite{15fo, 19gf}.
A comparison with the results for that of electrons would 
offer opportunities to extract the coefficients that control only CPT-odd effects
from the combination of coefficients in the difference \rf{gratio-lv}.
A sidereal-variation analysis of the positron anomaly frequencies
would offer limits on the starred tilde coefficients 
$\btws e J$, $\bftws e {(JK)}$
as well.

At the end of this subsection,
we note that from \Tab{pgconstraints} and \Tab{epconstraints},
for the 18 independent components of the tilde coefficients 
for Lorentz violation that are relevant to the $g$ factor measurements
in Penning-trap experiments,
only 6 of them in the proton sector and 12 of them in the electron sector
have been constrained so far.
Performing a full sidereal-variation analysis for the measurements data
would permit access to the other components of the tilde coefficients.

\section{Summary} 
\label{summary} 

In this work,
we provide an overview of recent progress on 
searching for Lorentz- and CPT-violating
signals using measurements of charge-to-mass ratios
and magnetic moments in Penning-trap experiments.
We first revisit the theory of Lorentz-violating quantum electrodynamics 
with operators of mass dimensions $d\leq 6$.
The explicit expressions of the Lagrange density \rf{fermlag} 
are given in \Refe{16dk}.
Perturbation theory is then applied to
obtain the dominant energy shifts \rf{linear-en} and \rf{linear-ens} 
for a confine particle and antiparticle due to Lorentz and CPT violation.
This leads to the corresponding contributions to the cyclotron 
and anomaly frequencies in Equations \rf{wcshift} and~\rf{washift}.
The results are then used to relate the coefficients for Lorentz violation to the
experimental interpreted charge-to-mass ratio comparisons \rf{ratio-lv}
between a particle and an antiparticle,  
as well as the magnetic moment comparisons~\rf{gratio-lv}.
The general transformation of the related coefficients
for Lorentz violation into different frames is performed using \Eq{axis}. 
The explicit expressions relating the coefficients in the apparatus frame
to the Sun-centered frame are given in Appendix \ref{transformations}.
The results show that for the charge-to-mass ratio comparisons 
between particles and antiparticles in Penning-trap experiments,
the related coefficients for Lorentz violation in the Sun-centered frame are 
$\bptw w {J}$,
$\ctw w {(JK)}$,
$\btw w {J(KL)}$,
$\bptws w {J}$,
$\ctws w {(JK)}$,
and
$\btws w {J(KL)}$.
For experiments involving magnetic moment comparisons,
the corresponding coefficients are
$\btw w J$, $\btws w J$, $\bftw w {(JK)}$, and~$\bftws w {(JK)}$.
Using published results from existing Penning-trap experiments, 
constraints on various components of the coefficients for Lorentz violation
are obtained.
They are summarized 
in Tables \ref{cons-p}-\ref{epconstraints}.
In conclusion,
the high-precision measurements and excellent coverage of the coefficients
for Lorentz violation offered by Penning-trap experiments provide
strong motivations to continue the searches for possible 
Lorentz- and CPT-violating signals.

\appendix
\section{Transformations}
\label{transformations}

In this Appendix,
we list the explicit expressions of the transformation results for the tilde coefficients for Lorentz violation
that are relevant to the cyclotron and anomaly frequency shifts. 
These coefficients are
$\bptw w {3}$, 
$\ctw w {11} + \ctw w {22}$,
$\btw w {311} + \btw w {322}$,
$\btw w {3}$,
$\bftw w {3}$,
$\bptws w {3}$, 
$\ctws w {11} + \ctws w {22}$,
$\btws w {311} + \btws w {322}$,
$\btws w {3}$,
and
$\bftws w {3}$
in the apparatus frame.

We first consider Penning-trap experiments that use a vertical upward magnetic field.
Taking $R^{aj}$ as the identity matrix and applying transformation \rf{axis} yield
\bea
\bptw w {3} &=& 
\codt  
\bptw w {X} \sin\ch 
+ \sodt
\bptw w {Y} \sin\ch 	
+ 
\bptw w {Z} \cos\ch ,
\eea

\bea
\ctw w {11} + \ctw w {22} &=& 
\ctodt  
\Big(  - \half (\ctw w {XX} -  \ctw w {YY}) \sin^2 \ch \Big)
+ \stodt
\left(  -  \ctw w {(XY)} \sin^2 \ch	 \right)
\nn \\
&& 
+  \codt  
\left(  - \ctw w {(XZ)}  \sin 2\ch \right)
+ \sodt
\left(  - \ctw w {(YZ)}  \sin 2\ch  \right)
\nn \\
&&
+ 
\frac{1}{4} (\ctw w {XX} + \ctw w {YY})(3+\cos 2\ch) + \ctw w {ZZ} \sin^2\ch  ,
\eea

\bea
\btw w {311} + \btw w {322} &=& 
\cthodt  
\left( [-\frac{1}{4} (\btw w {XXX} - \btw w {XYY}) + \half \btw w {Y(XY)}  ] \sin^3 \ch \right)
\nn \\
&&
+ \sthodt
\left( [- \half \btw w {X(XY)}  - \frac{1}{4} (\btw w {YXX} - \btw w {YYY}) ] \sin^3 \ch	\right)
\nn \\
&&
+\ctodt  
\left(  [ - \btw w {X(XZ)} + \btw w {Y(YZ)} - \half (\btw w {ZXX} - \btw w {ZYY}) ] \cos\ch \sin^2 \ch \right)
\nn \\
&&
+ \stodt
\left( [- \btw w {X(YZ)} - \btw w {Y(XZ)} - \btw w {Z(XY)} ] \cos\ch \sin^2 \ch \right)
\nn \\
&& 
+  \codt  
\Big( \frac{1}{8} \btw w {XXX} (5+3 \cos 2\ch)\sin\ch + \frac{1}{8} \btw w {XYY} (7+ \cos 2\ch)\sin\ch + \btw w {XZZ} \sin^3\ch 
\nn \\
&&
\hskip 60pt
- \frac{1}{2} \btw w {Y(XY)} \sin^3\ch - 2 \btw w {Z(XZ)} \cos^2\ch \sin\ch \Big)
\nn \\
&&
+ \sodt
\Big( - \frac{1}{2} \btw w {X(XY)} \sin^3\ch +  \frac{1}{8} \btw w {YXX} (7+ \cos 2\ch)\sin\ch +  \frac{1}{8} \btw w {YYY} (5+ 3\cos 2\ch)\sin\ch 
\nn \\
&&
\hskip 60pt
+  \btw w {YZZ} \sin^3\ch -  2\btw w {Z(YZ)} \cos^2\ch \sin\ch \Big)
\nn \\
&&
- (\btw w {X(XZ)} + \btw w {Y(YZ)} - \btw w {ZZZ} )\cos\ch \sin^2\ch
+ (\btw w {ZXX} + \btw w {ZYY}) \cos\ch \cos2\ch  ,
\eea

\bea
\btw w {3} &=& 
\codt  
\btw w {X} \sin\ch 
+ \sodt
\btw w {Y} \sin\ch 	
+ 
\btw w {Z} \cos\ch ,
\eea
and
\bea
\bftw w {33}
&=&
\cos 2\om_\oplus T_\oplus
\Big(
\half (\bftw w {XX} - \bftw w {YY}) \sin^2\ch \Big)
+ \sin 2\om_\oplus T_\oplus \bftw w {(XY)} \sin^2\ch \Big) 
\nn \\
&&
+\cos\om_\oplus T_\oplus
\bftw w {(XZ)} \sin 2\chi 
+ \sin\om_\oplus T_\oplus \bftw w {(YZ)} \sin 2\chi 
\nn \\
&&
+ \half (\bftw w {XX} +\bftw w {YY} -2\bftw w {ZZ}) \sin^2\ch + \bftw w {ZZ}.
\eea

We then switch to the case where a horizontal magnetic field is used in the trap,
directing at an angle $\th$ from the local south in the counterclockwise direction.
This implies that the Euler angles $(\al, \be, \ga) =  (\th, \pi/2, 0)$.
Applying the transformation~\rf{axis} we have
\bea
\bptw w {3} &=& 
\codt  
\left( \bptw w {X} \cos\th \cos\ch + \bptw w {Y} \sin\th \right)
+ \sodt
\left( - \bptw w {X} \sin\th + \bptw w {Y} \cos\th \cos\ch \right)
\nn \\
&&
- \bptw w {Z} \cos\th \sin\ch , 
\eea

\bea
\ctw w {11} + \ctw w {22} &=& 
\ctodt  
\left(  \frac{1}{8} (\ctw w {XX} -  \ctw w {YY}) (1-3\cos2\th - 2\cos^2\th \cos2\ch) - \ctw w {(XY)} \cos\ch \sin 2\th \right)
\nn \\
&& 
+ \stodt
\left(   \half (\ctw w {XX} -  \ctw w {YY}) \cos\ch \sin 2\th + \frac{1}{4} \ctw w {(XY)} (1-3\cos2\th - 2\cos^2\th \cos2\ch)	 \right)
\nn \\
&& 
+  \codt  
\left( \ctw w {(XZ)}  \cos^2\th \sin2\ch + \ctw w {(YZ)} \sin2\th \sin\ch \right)
\nn \\
&& 
+ \sodt
\left(  - \ctw w {(XZ)} \sin2\th \sin\ch + \ctw w {(YZ)}  \cos^2\th \sin2\ch  \right)
\nn \\
&&
+ 
\half(\ctw w {XX} + \ctw w {YY}) (\cos^2\th + \cos^2\ch \sin^2\th + \sin^2\ch ) + \ctw w {ZZ} (\cos^2\ch + \sin^2\th \sin^2\ch) ,
\eea

\bea
&&
\btw w {311} + \btw w {322}
\nn \\
 &=&
\cthodt  
\Big( [\frac{1}{64} (\btw w {XXX} -\btw w {XYY}) - \frac{1}{32} \btw w {Y(XY)}] [3 (\cos \theta -5 \cos 3 \theta ) \cos \chi -4 \cos ^3\theta  \cos 3 \chi ]
\nn \\
&&
\hskip 55pt
+ [\frac{1}{16}\btw w {X(XY)} + \frac{1}{32} (\btw w {YXX} -\btw w {YYY})] [3\sin \theta (1-4 \cos ^2 \theta  \cos 2 \chi) - 5 \sin 3 \theta ]  \Big)
\nn \\
&&
+ \sthodt
\Big( [- \frac{1}{32} (\btw w {XXX} - \btw w {XYY}) + \frac{1}{16} \btw w {Y(XY)}] [3\sin \theta (1-4 \cos ^2 \theta  \cos 2 \chi) - 5 \sin 3 \theta ]
\nn \\
&&
\hskip 60pt
+ [\frac{1}{32} \btw w {X(XY)} + \frac{1}{64} (\btw w {YXX} -\btw w {YYY})]  [3 (\cos \theta -5 \cos 3 \theta ) \cos \chi -4 \cos ^3\theta  \cos 3 \chi ] \Big)
\nn \\
&&
+\ctodt  
\Big( [\frac{1}{16} \btw w {X(XZ)} - \frac{1}{16}  \btw w {Y(YZ)} + \frac{1}{32} (\btw w {ZXX} -\btw w {ZYY})] [(5 \cos 3 \theta - \cos \theta \sin \chi +4 \cos ^3\theta  \sin 3 \chi ]
\nn \\
&&
\hskip 60pt
+(\btw w {X(YZ)} +\btw w {Y(XZ)} +\btw w {Z(XY)} )\sin \theta\cos^2\theta\sin 2\chi \Big)  
\nn \\
&&
+ \stodt
\Big( [-\btw w {X(XZ)} +\btw w {Y(YZ)} - \half (\btw w {ZXX} - \btw w {ZYY})] \sin \theta\cos^2\theta\sin 2\chi
\nn \\
&&
\hskip 60pt
- (\frac{1}{16} \btw w {X(YZ)} + \frac{1}{16} \btw w {Y(XZ)} + \frac{1}{16} \btw w {Z(XY)} ) [(\cos \theta -5 \cos 3 \theta ) \sin \chi -4 \cos ^3\theta  \sin 3 \chi ] \Big)
\nn \\
&& 
+  \codt  
\Big( \frac{1}{16} \btw w {XXX} \cos \theta \cos \chi (-6 \cos ^2 \theta \cos 2 \chi +3 \cos 2 \theta +7)
\nn \\
&&
\hskip 60pt 
- \frac{1}{2} \btw w {X(XY)}  \sin \theta (\cos ^2 \theta  \cos 2 \chi +\sin ^2 \theta  \cos ^2\chi +\sin ^2\chi )
\nn \\
&&
\hskip 60pt 
+ \frac{1}{16} \btw w {XYY} \cos \theta\cos \chi(-2 \cos^2\theta\cos 2\chi +\cos 2\theta+13)
\nn \\
&&
\hskip 60pt 
+\btw w {XZZ} \cos \theta\cos \chi(\sin^2\theta\sin^2\chi+\cos^2\chi)
\nn \\
&&
\hskip 60pt
+ \frac{1}{32} \btw w {YXX} [\sin \theta(25-4 \cos^2\theta\cos 2\chi )+\sin 3 \theta]\nn \\
&&
\hskip 60pt 
+ \frac{1}{16} \btw w {Y(XY)} \cos \theta[(\cos 2\theta-7) \cos \chi-2 \cos^2\theta\cos 3 \chi]
\nn \\
&&
\hskip 60pt
+ \frac{1}{32} \btw w {YYY} [\sin \theta(11-12 \cos^2\theta\cos 2\chi )+3 \sin 3 \theta]
\nn \\
&&
\hskip 60pt
+\btw w {YZZ} \sin \theta(\sin^2\theta\sin^2\chi+\cos^2\chi)
\nn \\
&&
\hskip 60pt
-2\btw w {Z(XZ)} \cos^3\theta\sin^2\chi\cos \chi
-2\btw w {Z(YZ)} \sin \theta\cos^2\theta\sin^2\chi \Big)
\nn \\
&&
+ \sodt
\Big(\frac{1}{32} \btw w {XXX} [\sin \theta(12 \cos^2\theta\cos 2\chi -11)-3 \sin 3 \theta]
\nn \\
&&
\hskip 60pt 
+ \frac{1}{16} \btw w {X(XY)} \cos \theta[(\cos 2\theta-7) \cos \chi-2 \cos^2\theta\cos 3 \chi]
\nn \\
&&
\hskip 60pt
+ \frac{1}{32} \btw w {XYY} [\sin \theta(4 \cos^2\theta\cos 2\chi -25)-\sin 3 \theta]
\nn \\
&&
\hskip 60pt 
-\btw w {XZZ} \sin \theta(\sin^2\theta\sin^2\chi+\cos^2\chi)
\nn \\
&&
\hskip 60pt
+\frac{1}{16} \btw w {YXX} \cos \theta\cos \chi(-2 \cos^2\theta\cos 2\chi +\cos 2\theta+13)
\nn \\
&&
\hskip 60pt 
+ \frac{1}{2} \btw w {Y(XY)} \sin \theta (\cos^2\theta\cos 2\chi +\sin^2\theta\cos^2\chi+\sin^2\chi)
\nn \\
&&
\hskip 60pt
+ \frac{1}{16} \btw w {YYY} \cos \theta\cos \chi(-6 \cos^2\theta\cos 2\chi +3 \cos 2\theta+7)
\nn \\
&&
\hskip 60pt 
+\btw w {YZZ} \cos \theta\cos \chi(\sin^2\theta\sin^2\chi+\cos^2\chi)
\nn \\
&&
\hskip 60pt
+2\btw w {Z(XZ)} \sin \theta\cos^2\theta\sin^2\chi
-2\btw w {Z(YZ)} \cos^3\theta\sin^2\chi\cos \chi \Big)
\nn \\
&&
+ 
\frac{1}{8} (\btw w {X(XZ)} + \btw w {Y(YZ)} - \btw w {ZZZ}) \cos \theta (-4 \cos 2\theta \sin^3 \chi + 5 \sin \chi + \sin 3\chi)
\nn \\
&&
+ \frac{1}{16} (\btw w {ZXX} + \btw w {ZYY}) [2 \cos^3 \theta \sin 3\chi - \cos \theta \sin \chi (3 \cos 2\theta + 11 )] ,
\eea

\bea
\btw w {3} &=& 
\codt  
\left( \btw w {X} \cos\th \cos\ch + \btw w {Y} \sin\th \right)
+ \sodt
\left( - \btw w {X} \sin\th + \btw w {Y} \cos\th \cos\ch \right)
\nn \\
&&
- \btw w {Z} \cos\th \sin\ch , 
\eea
and 
\bea
\bftw w {33}
&=& 
\ctodt  
\Big(  \half (\bftw w {XX} -  \bftw w {YY}) (\cos^2 \th \cos^2 \ch - \sin^2 \th) +  \bftw w {(XY)} \sin 2\th \cos\ch \Big)
\nn \\
&& 
+ \stodt
\Big(  \frac{1}{4} \bftw w {(XY)} (2\cos^2\th \cos2\ch+3\cos2\th - 1)- (\bftw w {XX} -  \bftw w {YY}) \sin2\th \cos\ch \Big)
\nn \\
&& 
+  \codt  
\Big(  -2\bftw w {(XZ)}  \cos^2\th \cos\ch - \bftw w {(YZ)} \sin2\th  \sin\ch \Big)
\nn \\
&& 
+ \sodt
\Big(  \bftw w {(XZ)} \sin2\th  \sin\ch - 2\bftw w {(YZ)}  \cos^2\th \cos\ch  \Big)
\nn \\
&&
+ 
\frac{1}{2} (\bftw w {XX} + \bftw w {YY})(\cos^2 \th \cos^2\ch +\sin^2\th) + \bftw w {ZZ} \cos^2\th \sin^2\ch .
\eea 

The corresponding transformations for the starred tilde quantities 
$\bptws w {3}$, 
$\ctws w {11} + \ctws w {22}$,
$\btws w {311} + \btws w {322}$,
$\btws w {3}$,
and 
$\bftw w {33}$
take the same forms as these given above,
with the substitutions of the usual tilde coefficients by the starred ones.

\vspace{8pt}

\funding{
This work was supported in part by the
Gettysburg College Research and Professional Development Grant,
the Gettysburg College Cross-Disciplinary Science Institute (X-SIG),
and the Keck Grant from the W.M. Keck Science Department at
Claremont McKenna, Pitzer, and Scripps Colleges.}

\acknowledgments{
The authors would like to thank Matthew Mewes for the invitation and Lisa Portmess for reading the manuscript.} 

\conflictsofinterest{The authors declare no conflict of~interest.} 

\appendixtitles{yes} 
\appendixsections{one} 



\reftitle{References}



\end{document}